\documentclass[
    ,final            
  ]
  {aipproc}

\layoutstyle{6x9}

\begin{document}

\title{X\&Y}

\author{L.~Maiani}{
address={Universit\`{a} di Roma `La Sapienza' and INFN Sezione di Roma, Italy}
}
\author{F.~Piccinini}{
  address={INFN Sezione di Pavia and Dipartimento di Fisica Nucleare e Teorica, Pavia, Italy}
}
\author{A.D.~Polosa\footnote{Speaker}}{
  address={Dip. Interateneo di Fisica, Universit\`{a} di Bari, Italy}
}
\author{V.~Riquer}{
  address={INFN Sezione di Roma, Italy}
}

%

\begin{abstract}
The 4-quark interpretation of the recently discovered 
$X(3872)$ and $Y(4260)$ resonances is briefly discussed.  
\end{abstract}

\maketitle


\section{ $X(3872)$}

Quite recently BELLE and BaBar (also CDF and D0) have reported evidence of a new narrow resonance in $B^+\to K^+ J/\psi \pi\pi(\pi)$, the $X(3872)$~\cite{xx}, which most likely decays to $X\to J/\psi\rho(\omega)$
with almost equal rate, therefore maximally violating isospin. Even if its mass is right at the $D\bar{D}$ threshold, the X(3872) is not observed to decay in this channel. It turns out that it can't because $J^P(X)=1^+$. The decay $X\to J/\psi \gamma$, with a rate almost ten times smaller than the $J/\psi\rho$,  is observed, suggesting $C=+1$, i.e., $J^{PC}(X)=1^{++}$.

\paragraph{Charmonium and Hybrid}
Such a state is a natural candidate for being classified as a new $c\bar c$ resonance. Anyway, the experimental assessment of $X$ is in sharp contrast with charmonium assignments~\cite{barnes}.
What is this narrow $X$ then? The hypothesis next to the charmonium one is the hybrid, i.e., a 
$c\bar c+$constituent$(g)$.
Anyway the gluon is flavor blind, it couples to $u$ and $d$ quarks in the same way thus making the hybrid state an isospin pure state. Isospin violating decays forbid this scheme.

\paragraph{Molecule}
There are at least two other possibilities for the $X$: being a $D\bar{D}^*$ molecule (a $4\div 5$~fm extended object)~\cite{torn} or a multiquark state (a compact 4-quark state with a size of$\sim$1~fm)~\cite{noi}. The molecule is a $50\%$ mixture of $I=0$ and $I=1$, allowing isospin violating transitions. The $D^*$ has a width of about $70$~KeV. The width of such a molecule is expected to be due to the $D^*$ width. A quite natural decay is then $DD\pi$, which is currently searched for at BELLE and BaBar~\cite{pak}. Preliminary  results agree that this channel is at least ${\cal O}(1)$ with respect to the $J/\psi\rho$ one. On the other hand $D$ and $D^*$ might as well interact via hadron exchange forces giving $J/\psi \rho$ in the final state. In more sophisticated models the $X$ spends part of its time in a molecular-like state and part in a compact multiquark-like state; once the $c$ quarks are closer to each other, the  transition to a final state containing a $J/\psi$ is favored. 

\paragraph{4-quark}
In the four-quark option the $X$ is depicted as a bound state of two colored diquarks $[cq]$ and
$[\bar{c}\bar{q}]$. The heavy-light diquarks are supposed to be bound in the $\bar 3_c$ attractive channel~\cite{noi, jw}. 
There is some lattice evidence of the formation of $S=0$ light-light diquarks, while $S=1$ 
seems to be disfavored~\cite{lucini}. In the heavy-light system, spin-spin interactions are suppressed by the mass of the heavy constituent, therefore we expect that, {\it if} the heavy-light diquark
is formed,  $S=1$ or $S=0$ will be equally probable. This opens the way to 4-quark mesons with $J=0,1,2$ and natural and unnatural parities.
Depending upon the flavor of the two light constituents, one can build a full nonet of  $X$ states where the observed $X(3872)$ is just the neutral one (charged states are expected to occur!).

Diquarks are colored objects. If one tries to pull them apart, at some point will excite the production of a couple of light quarks from the vacuum. This means that the natural decay of such a four quark system is into two charmed baryons. But this is impossible kinematically. Thus some other mechanism must be at work to allow the $X$ decay.
If the two heavy $c$ quarks escape from their diquark shells and bind to form a $J/\psi$, then a $X\to J/\psi \rho(\omega)$ decay is easily explained. Also the two diquark structures could be broken by the tunneling of a light anti-quark from the anti-diquark to  meet a $c$ quark in the diquark (and viceversa) then forming a $D \bar D^*\to DD\pi$ final state. 

Isospin forbidden decays are possible if $X$ is not a pure isospin state. A pure $I=0,1$ state would be:
\begin{equation}
X(I=0)=\frac{X_u+X_d}{\sqrt{2}} \;\;\;{\rm or}\;\;\; X(I=1)=\frac{X_u-X_d}{\sqrt{2}}\nonumber,
\end{equation}
where $X_{u,d}=[cu][\bar c \bar u], [c d][\bar c\bar d]$. On the contrary if the physical eigenstates are 
just  $X_u$ and $X_d$, i.e., pure mass eigenstates, maximal isospin violations are possible.
Mixtures of $X_u$ and $X_d$ are induced by annihilation diagrams where $u$ quarks transform in $d$ through vacuum. At the scale of the charm quark mass $m_c$, we expect that annihilation diagrams 
are suppressed enough to be close to the situation in which $X$ mass eigenstates tend to align to quark masses. We can introduce a $\theta$ mixing angle and write~\cite{noi}:
\begin{eqnarray}
X_{l}&=&X_u\cos \theta +X_d\sin \theta \nonumber\\
X_{h}&=&-X_u\sin \theta +X_d\cos \theta, \nonumber
\end{eqnarray}
expecting  $\theta$ to be small.  
We find that the mass of $X_h$ and $X_l$ must be different by:
\begin{equation}
M(X_h)-M(X_l)\simeq \frac{2(m_d-m_u)}{\cos (2\theta)}\nonumber
\end{equation}
and our estimate is $\Delta M\simeq 7$~MeV.

Taking Belle data at face value, we conclude that only one of the two neutral states, $X_{u,d}$ or, more generally $X_{l,h}$, is  produced appreciably in $B^+$ decays: the peak observed is indeed too narrow to describe two resonances about 7~MeV apart. Only one should be produced; our analysis is consistent with the production of $X_l$. $X_h$ should then be produced in $B^0$ decays.
The present BaBar measure of this mass difference is $\Delta M=2.7\pm1.3\pm 0.2$~MeV, extracted from $B^0\to X K^0$ and $B^+\to X K^+$~\cite{babdiff}. We have to wait for higher statistics analyses to clarify if there are indeed two neutral states separated in mass.
\begin{figure}
\includegraphics[height=.3\textheight]{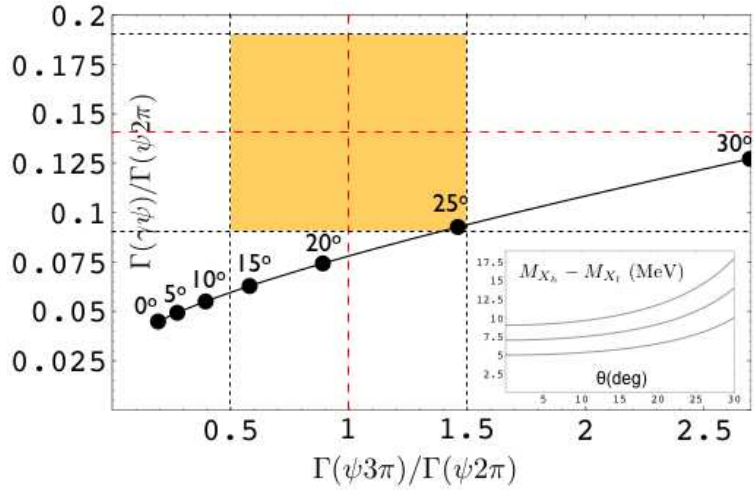}
\caption{The shaded area represents the experimentally allowed region by the decays $X\to J/\psi+ \gamma,\rho,\omega$. The solid curve is the 4-quark prediction for $X_l$ at different values of $\theta$. The insert shows the variation of the mass difference $M_{X_h}-M_{X_l}$ versus $\theta$ including the theoretical error band. }
\end{figure}

\paragraph{Differences}
 In Fig.~1 the $X_l$ widths as a function of $\theta$ in the 4-quark model. Data seem to favor a $\theta\sim 20^\circ$.  
Let's summarize some of the more striking differences between 4-quark models and molecules.
The four quark model predicts the existence of a full nonet of $X$ states. Charged $X$ particles are predicted. Two neutral $X$ particles are predicted with a separation in mass. In this framework the $X(3940)$, seen in $J/\psi \omega$ by BELLE~\cite{abe}, could be interpreted as a $2^{++}$ 4-quark state. Does it decay also to $J/\psi \rho$? 

On the other hand the molecule approach to $X$ does not call for charged $X$'s, $X_s$'s or for two distinct neutral $X$'s. The 4-quark model for $X$ allows 
a $X\to D\bar D^*$ decay width up to two times larger than $J/\psi \rho$. The molecule cannot, 
the $D\bar D^*$ being quite smaller than $J/\psi \rho$ (see e.g.~\cite{swa}).
Experiment will shortly decide between (or beyond) these two schemes.

\section{$Y(4260)$}

Colored objects such as diquarks in a rising confining potential should exhibit 
a series of orbital angular momentum excitations while in the
molecular picture two colorless objects bound by a short range potential 
should have a very limited spectrum, possibly restricted to  S-wave states only.

The $Y(4260)$ has been observed by BaBar as a $1^{--}$ resonance in $e^+e^-\to\gamma J/\psi \pi\pi$, $\gamma$ being an initial state radiation photon.  The two pion invariant mass spectrum seems to be peaked at the $f_0(980)$ (this awaits experimental confirmation). $Y(4260)$ is not observed to decay into $D\bar D$. Is it the first excited state of the $X$ family?~\cite{noi2}

\paragraph{4-quark}
We reconstruct the process as $e^+e^-\to \gamma Y\to \gamma J/\psi f_0 $.  Negative parity of the state calls for one unit of orbital angular momentum in the diquark pair. The final state $f_0+J/\psi$ suggests, in the 4-quark model of $f_0$~\cite{scal4q}, the quark content $Y\sim \{ [cs][\bar c \bar s]\}_{\rm (P-wave)}$.
The most striking prediction of the 4-quark approach is that the observed decay  $J/\psi f_0$ should not be the favored one. The most natural decay should be $D_s \bar D_s$, as  shown in Fig.~2.
 
\begin{figure}
\includegraphics[height=.15\textheight]{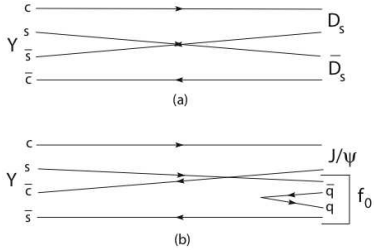}
\caption{Diagram (a) is favored with respect to (b).}
\end{figure}

The 4-quark model predicts a series of states in addition to the observed one.
Among them, a state with both diquark and anti-diquark with $S=1$ can give rise to a $S_{\rm tot}=2$ state. The latter projects only on spin one $\bar c s$ and $\bar s c$ states. In the limit in which the spin of the $s$ quark is a good quantum number such state could decay only into $D^*_s\bar D^*_s$ with substantial reduction of its decay width. One could then expect another state with higher mass, by a factor typical of spin-spin interactions, decaying predominantly to $D^*_s\bar D^*_s$.
 
\paragraph{Hybrid}
A serious competitor here is the hybrid state $c\bar c g$. This is predicted to be subject to selection rules constraining it to decay to $D^{**}\bar D$, all other modes being suppressed~\cite{kou}.
Also in this case, like in the case of X, a quite stringent experimental test to 
select among models (or disprove them all) is available.

\section{$B\to J/\psi+{\rm all}$}
The inclusive $J/\psi$ spectrum in $B$ decays has been studied very carefully on the experimental and theoretical side. The Non-Relativistic-QCD predictions agree quite well with the data over most of the range. However for slow  $J/\psi$, i.e., with momenta below 1~GeV, the data show a marked excess over expectations, see Fig~3. No definitive explanation has been furnished yet to justify this excess.

We propose that these soft $J/\psi$'s are the footprints of the production of hidden charm four-quark states in $b\to c\bar c q$ decays and their subsequent decay into $J/\psi$ plus light-flavor hadrons~\cite{noi3}. Whatever $X$ and $Y$ are made of, they could be present in a number of $B$ decays.

\begin{figure}
\includegraphics[height=.3\textheight]{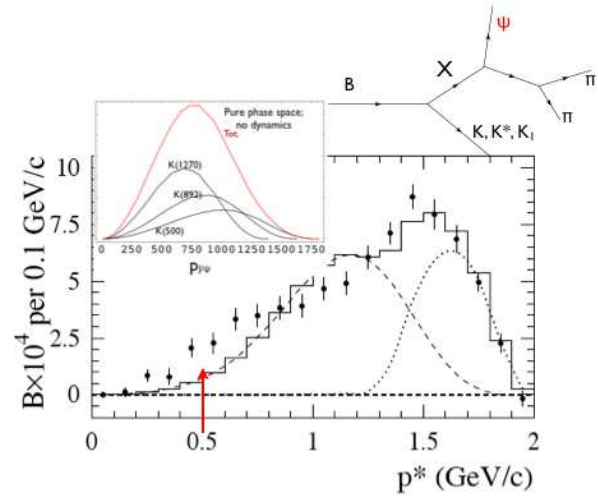}
\caption{The process $B\to J/\psi+{\rm all}$ could be dominated by a resonant channel in which the $J/\psi$ comes from an $X$ produced in association to a Kaon, P-wave, or to a $K^*$  or $K_1$, S-wave. Pure kinematics considerations immediately show that the $J/\psi$ decay momenta should be most likely in the  low region, especially when a $K^*$ or $K_1$ is produced (see the inserted plot). This would justify the $J/\psi$ momentum spectrum (main plot) where a clear discrepancy with non-relativistic QCD is appreciable at low $p_{J/\psi}$ (the region around 500~MeV).}
\end{figure}

\section{Summary}

$X$ and $Y$ particles, and their (coming...) partners, could be the first pieces of a new series of mesons with new multiquark body plans (or hybrids?) or could be the first examples of loosely bound states of charmed mesons in the form of molecules. 
There are several crucial experimental tests which will allow to probe the nature of this states and distinguish among models.





\begin{theacknowledgments}
ADP thanks R.~Faccini and E.~Robutti for many informative discussions.
\end{theacknowledgments}



\bibliographystyle{aipproc}   

\bibliography{sample}




\end{document}